\documentclass[twocolumn]{article} 
\usepackage{amsmath,amssymb,amsfonts}
\usepackage{simpleConference}

\usepackage{booktabs} 

\usepackage{booktabs} 
\usepackage{graphicx}
\usepackage[title]{appendix}
\usepackage{kotex}
\usepackage{wrapfig}
\usepackage{multirow}
\usepackage{float}
\usepackage{amsmath}

\DeclareMathOperator*{\argmin}{arg\,min}

\usepackage{tabularx}
\usepackage{algpseudocode}
\usepackage{float}
  \floatstyle{ruled}
  \newfloat{algorithm}{thp}{alg}
\floatname{algorithm}{Algorithm}
\usepackage{algorithmicx}
\usepackage{authblk}

\title{Generative GaitNet}

\author[1]{Jungnam Park\thanks{jungnam04@mrl.snu.ac.kr}}
\author[1]{Sehee Min\thanks{sehee@mrl.snu.ac.kr}}
\author[1]{Phil Sik Chang\thanks{phil@mrl.snu.ac.kr}}
\author[1]{Jaedong Lee \thanks{Jaedong@mrl.snu.ac.kr}}
\author[2]{Moonseok Park\thanks{pmsmed@gmail.com}}

\author[1]{Jehee Lee \thanks{jehee@mrl.snu.ac.kr}}
\affil[1]{Seoul National University, Seoul, South Korea}
\affil[2]{Seoul National University Bundang Hospital}
\date{ }

\begin{document}

\maketitle

\begin{abstract}
Understanding the relation between anatomy and gait is key to successful predictive gait simulation. In this paper, we present Generative GaitNet, which is a novel network architecture based on deep reinforcement learning for controlling a comprehensive, full-body, musculoskeletal model with 304 Hill-type musculotendons. The Generative Gait is a pre-trained, integrated system of artificial neural networks learned in a 618-dimensional continuous domain of anatomy conditions (e.g., mass distribution, body proportion, bone deformity, and muscle deficits) and gait conditions (e.g., stride and cadence). The pre-trained GaitNet takes anatomy and gait conditions as input and generates a series of gait cycles appropriate to the conditions through physics-based simulation. We will demonstrate the efficacy and expressive power of Generative GaitNet to generate a variety of healthy and pathologic human gaits in real-time physics-based simulation. 
\end{abstract}

\section{Introduction}

A long-standing goal of human animation is a simulation algorithm that predicts gait in specific conditions. Specifically, we consider predictive gait simulation that takes anatomical conditions (e.g., mass distribution, body proportion, bone deformity, and muscle weakness/contracture) and gait conditions (e.g., stride and cadence) as input and generates a series of periodic gait cycles appropriate to the given conditions via physics-based simulation. This goal with a musculoskeletal simulation model is challenging owing to its large control space and the difficulty of evaluating the combined dynamics of muscle contraction and bone articulations. Feedback/feedforward control, linear/nonlinear control theory, various forms of trajectory optimization, anatomical modeling and simulation have been studied to achieve robust gait simulation for several decades~\cite{wang2012optimizing, geijtenbeek2013flexible, lee2014locomotion}. Recently, rapid progress toward this goal has been made using DRL (deep reinforcement learning)~\cite{kidzinski2018learning, lee2019scalable}. 

The core of predictive gait simulation is establishing a conceptual mapping between human anatomy and its gait. Ideally, we wish that appropriate gait emerges from biological principles for any given body condition. This approach often requires dynamical models much simpler than human anatomy~\cite{kajita2003biped, wang2012optimizing} or results in aberrant gaits that do not look human-like~\cite{heess2017emergence,waterval2021validation}. Alternatively, many gait simulation algorithms leverage motion capture data as reference to formulate the problem as trajectory tracking~\cite{lee2014locomotion, peng2018deepmimic, fussell2021supertrack}. Given a reference trajectory to track, gait simulation algorithms can faithfully reproduce every nuance of human gaits captured in the reference data and the simulation can also be more robust and stable. Despite the advantages of reference-tracking approaches, a reference trajectory is considered harmful for predictive gait simulation because the simulated gait is constrained by the reference trajectory and thus may not fully adapt to new conditions as desired. 

Although the state-of-the-art algorithms based DRL have demonstrated highly-detailed, anatomically-valid gaits, the process is computationally demanding. Given a reference trajectory, muscle-actuated gaits can be reproduced for a specific anatomy condition through policy learning and physics-based simulation. Policy learning takes several hours to several days depending on the complexity of the anatomical model and the level of dynamic skills. Whenever new conditions are given, a new policy must be learned from scratch.

In this paper, we present a Generative GaitNet (or simply GaitNet), which is a pre-trained, integrated system of artificial neural networks that learns the control policy of human gaits eligible for a high-dimensional continuous domain of anatomical conditions. The control policy is massively parameterized by a large number of bone and muscle conditions. The GaitNet produces a broad spectrum of human gaits starting from a single motion clip of the reference gait cycle, which serves as gentle guidance of cyclic patterns. Accurately tracking the reference motion is not the primary goal of the GaitNet. With the GaitNet, simulated gaits can deviate from the reference motion by using biologically-inspired minimal rewards instead of imitation rewards and employing hierarchical motion displacement mapping and variable phase stepping. The key to the success of Generative GaitNet is our novel CSN (Cascaded Subsumption Network) architecture that captures the correlations among muscle activations and their influence on the simulated gait in a hierarchical manner. The CSN allows us to learn the large integrated policies progressively while alleviating the risks of catastrophic forgetting. 

We will demonstrate the efficacy and expressive power of Generative GaitNet to generate a broad variety of healthy and pathologic human gaits. The pre-trained GaitNet allows gait simulation to be interactive and responsive. The change of any anatomical condition is immediately reflected to the real-time simulation. 

\section{Related Work}

Reproducing human locomotion in physics-based simulation has been studied in graphics, robotics, and biomechanics for decades. Early approaches leveraged finite state machines and hand-crafted feedback rules to control dynamics systems~\cite{hodgins1995animating, yin2007simbicon}. Conceptual (often Simplified) dynamics models, such as an inverted pendulum and its variants, have been comprehensively explored to simulate the dynamics of biped locomotion~\cite{coros2010generalized, mordatch2010robust, kajita2003biped}. The use of motion capture data significantly improved the quality and stylistic variation of locomotion simulation~\cite{sok2007simulating,lee2010data}. Statistical optimal control theory provides a solid theoretical foundation for understanding the relation among energy efficiency, biological plausibility, and task achievement~\cite{yin2008continuation, wang2009optimizing}. Many trajectory optimization methods that minimize either total joint torques or metabolic energy expenditure have been explored~\cite{anderson2001dynamic, al2012trajectory,tassa2012synthesis,mordatch2014combining,kwon2017momentum, kwon2020fast}.

Recently, remarkable progresses have been made in biped locomotion control using DRL (Deep Reinforcement Learning), which allows robust control policies to be learned by deep artificial neural networks. It has been demonstrated that deep control policies can deal with uneven terrain and various motor skills~\cite{peng2016terrain,peng2017deeploco}. Deep control policies can imitate human motion closely if reference motion capture data are provided~\cite{peng2018deepmimic,park2019learning,won2020scalable,bergamin2019drecon, liu2018learning}. Although the use of reference motion data is often preferred for best-looking results, the reference motion is not mandatory. Stylistic, physically-plausible motions can be produced without any reference data~\cite{yu2018learning,merel2020catch}. The usability and flexibility of DRL are significantly improved if the control policies can be learned in continuously parameterized environments~\cite{portelas2020teacher,won2019learning,lee2021learning}. Unfortunately, previous studies reported that brute-force curriculum learning can deal with only a few parameters (up to four dimensions as reported by Lee et al.~\cite{lee2021learning}). Learning control policies in higher-dimensional parametric domain is a technical challenge we address in this work.


The dynamical model of the human body in aforementioned studies consists of rigid links connected by idealized motors. The anatomical simulation model consisting of bones, musculotendons, and their contraction dynamics achieves better accuracy, flexibility, and applicability in gait simulation. The combination of muscle actuation models and nonlinear optimization has successfully demonstrated the simulation of healthy and pathological human gaits~\cite{wang2012optimizing,thatte2015toward,peng2017learning,song2018predictive,ong2019predicting,waterval2021validation}. Since control optimization is notoriously time-consuming, early studies used simplified musculoskeletal models with only one or two dozens of musculotendon units. There have been continuing effort to develop efficient trajectory optimization and muscle approximation methods that can deal with scalable anatomical models with more muscles~\cite{lee2014locomotion,falisse2019rapid, dembia2020opensim,jiang2019synthesis}. In addition to humans, the locomotion of various creatures and underwater animals has also been studied~\cite{geijtenbeek2013flexible,min2019softcon}.

DRL has also brought great progress in muscle-actuated simulation and control~\cite{anand2019deep, song2020deep, wang2019terrain}. Many results of series of the NeurIPS AI for prosthetics challenge  produced muscle-actuated gaits of interactively-controllable speed~\cite{kidzinski2018learning, kidzinski2020artificial}. The state-of-the-art, DRL-based algorithm presented by Lee and his colleagues~\cite{lee2019scalable} uses a comprehensive anatomical model with 346 muscles. Their algorithm is capable of learning robust control policies that produce various human movements, such as running, jumping, and cartwheel, by mimicking reference motion data. Our Generative GaitNet is also based on DRL focusing on bipedal locomotion under various body sizes, proportions, and muscle deficits.

\section{Parameterized Anatomical Model}

\begin{figure}
    \centering
    \includegraphics[width=0.9\linewidth]{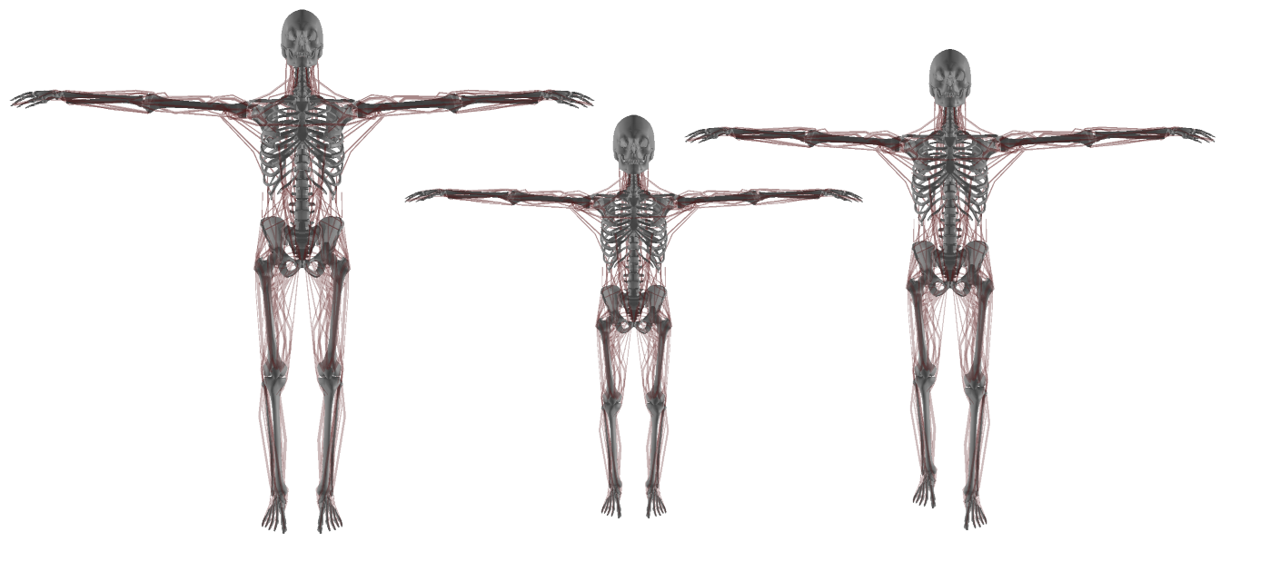}
    \caption{\label{fig:various_model} Parameterized musculoskeletal models. (Left) Reference Model. (Middle) Child. (Right) Uneven leg length.}
\end{figure}

Our anatomical model has 23 rigid bones and 304 muscles. Each muscle is attached to bones at its origin and insertion points. The geometric route of the muscle is represented by a series of waypoints, which transmit force between the origin and insertion points. The muscle generates contraction force according to Hill-type muscle model.
\begin{equation} \label{eq:hill-muslce} \centering
f_\mathrm{muscle}(l,\dot{l}, a) = f_\mathrm{max} \big(a  \cdot g_\mathrm{c} (l, \dot{l}) + g_\mathrm{p}(l) \big),
\end{equation}
where  $f_{\rm max}$ is the maximal isometric force. The muscle can generate its maximal force when its length is optimal $l_\mathrm{optimal}$. $l$ and $\dot l$, respectively, denote the muscle length and the rate of length change normalized by the optimal length. $a\in[0,1]$ is its activation level. $g_\mathrm{c}$ is the contractile force, which depends on the force-length and force-velocity curves. $g_\mathrm{p}$ is the passive force. The active force occurs when the muscle is active, while the passive force occurs when the muscle is stretched. We determined the parameters of the Hill-type model based on data sheets provided in previous studies~\cite{lee2019scalable,delp2007opensim}.

Our anatomical model is morphable with adjustable body and muscle conditions (see Figure~\ref{fig:various_model}). The reference model represents an healthy individual who is 168.7 cm tall and weighs 72.9 kg.
The skeleton is parameterized by ten parameters
\begin{equation}
    C_\mathrm{body} = (c_\mathrm{head}, c_\mathrm{torso}, c_1, \cdots, c_8)
\end{equation}
that generates a continuous spectrum of body sizes and proportions. $c_\mathrm{head}$ and $c_\mathrm{torso}$ are scale factors of the head and the torso, respectively, and the others are scale factors of four (upper and lower) limbs. The size, length, and weight of individual body parts are scaled accordingly. We consider two types of muscle deficits: weakness and contracture. The maximal isometric force indicates how strong the muscle is. The weakness parameter $c_\mathrm{weakness} \in [0,1]$ of a muscle is the scaling factor of its maximal isometric force with respect to the reference model. Contracture refers to permanent shortening of muscles and tendons. The contracture parameter $c_\mathrm{contracture} \in [0,1]$ of a muscle is the scaling factor of its optimal length. The anatomical condition of a parametrically-varied model is represented by a vector $\mathbf{C}_\mathrm{anatomy} = (C_\mathrm{body}, C_\mathrm{weakness}, C_\mathrm{contracture}) \in \mathbb{R}^{10+2\times 304}$ concatenating the body scale, weakness, and contracture factors over all body parts and muscles.

Given the size and proportion of the skeleton altered, the geometric route of each muscle and its parameters must be altered as well to match the change of the skeleton. We use the Musculature Retargeting algorithm by Ryu et al.~\cite{ryu2021functionality}, which systematically updates Hill-type muscle parameters and waypoints such that the muscle functionality and the range of joint motion are preserved for parameterically-varied models.

\section{Generative GaitNet} \label{sec:learntowalk}

\begin{figure}
    \centering
    \includegraphics[width=\linewidth]{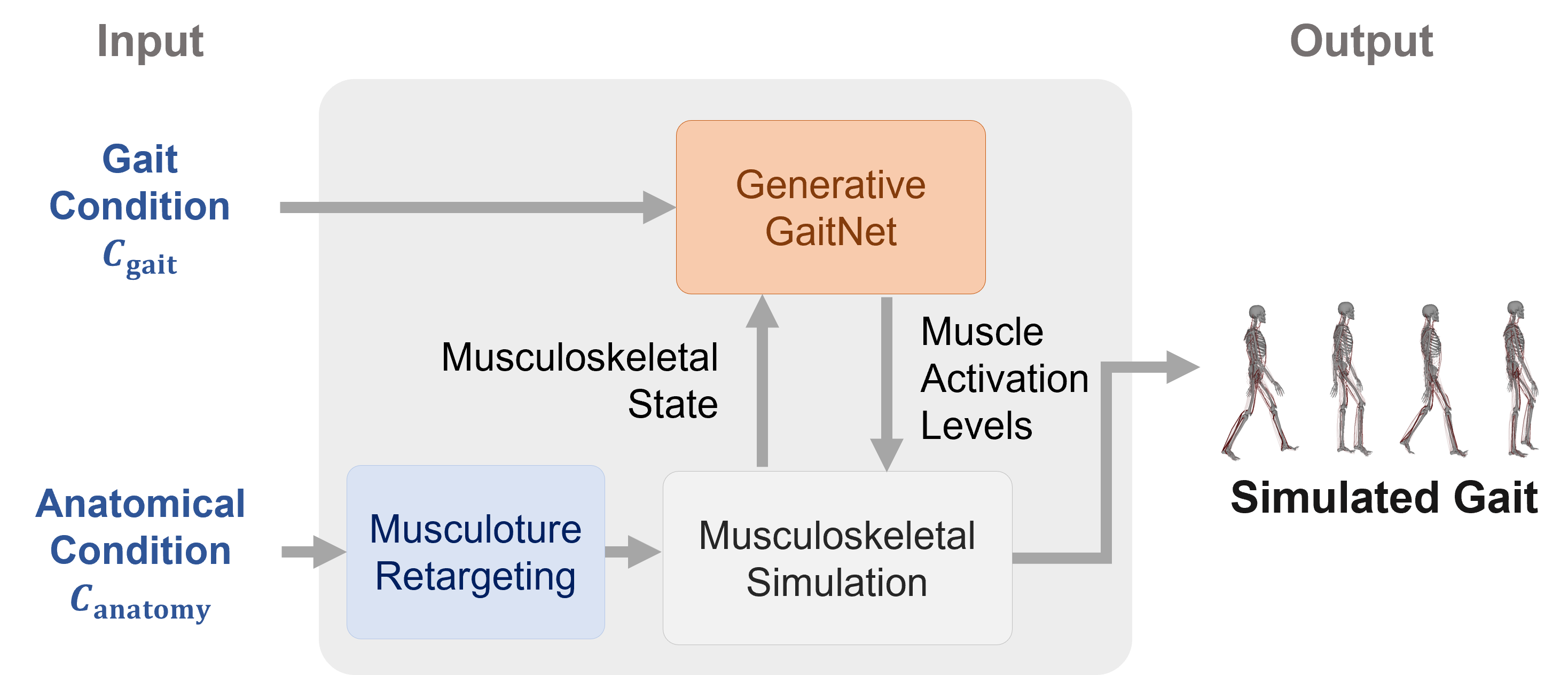}
    \caption{\label{fig:Overview} System Overview.}
\end{figure}

The Generative GaitNet is an integrated system of artificial neural networks that learns the relation between anatomy and gait (see Figure~\ref{fig:Overview}). The GaitNet takes gait condition $\mathbf{C}_\mathrm{gait}$ and anatomical condition $\mathbf{C}_\mathrm{anatomy}$ as input and generates the corresponding physically- and anatomically-plausible gait through physics-based musculoskeletal simulation. The gait condition include stride and cadence. High-dimensionality of the anatomy condition explained in the previous section makes the learning of GaitNet very challenging. In this section, we will explain a baseline algorithm, which works well with a moderate subset of the anatomical condition. We will explain how to learn the GaitNet with the full set of 608-dimensional anatomical condition through cascading and subsumption in the next section.

In deep reinforcement learning, an agent in state $s$ receives reward $r$ from the environment through action $u$. Through the action sequence of the agent in the environment, the cumulative reward $r_0 + \gamma r_1 + \gamma^2 r_2 + ... + \gamma^n r_\textrm{n}$ is calculated, where $\gamma$ is a discount factor. The goal of reinforcement learning is to learn the policy $\pi_\theta (u|s)$ that maximizes the cumulative reward. Deep reinforcement learning showed remarkable performance in torque control of 3D characters in physics-based simulation~\cite{peng2018deepmimic,park2019learning,won2020scalable}. Recently, DRL showed its potential in muscle-driven control using two-level hierarchical learning in the work of Lee et al~\cite{lee2019scalable}. We adopted the two-level network architecture, though our work is fundamentally different from their trajectory-tracking approach. Our anatomically-inspired rewards and variable phase-stepping method achieves realistic human gait without imitation rewards. Although we use a reference gait to guide the progression of the simulation, the simulation is not hindered by the guidance but is free to deviate from the reference gait to satisfy required conditions.

In the two-level network architecture, the torque-based controller $\pi_\theta(u|s)$ is a stochastic policy that produces PD target poses as output. The PD servos compute desired joint torques $\tau$ from the target poses. The muscle coordinator $a=\pi_\psi(\tau,s)$ is a regression network that computes muscle activation $a$ that drives the musculoskeletal system in physics-based simulator. The policy network $\pi_\theta$ and the regression network $\pi_\psi$ are learned jointly in the DRL framework.

\subsection{State and Action}

The state of the musculoskeletal system has three terms.
\begin{equation}
    s = (s_\mathrm{skeleton}, s_\mathrm{muscle}, s_\mathrm{joint}, s_\mathrm{gait}),
\end{equation}
where $s_\mathrm{skeleton}$ includes the center of mass (COM) velocity, the relative position and linear velocity of body parts with respect to COM, and their orientation and angular velocity. $s_\mathrm{muscle}$ includes muscle weakness and contracture parameters.

Multiple muscles around a joint are responsible for actuating the joint. This relation is denoted by
\begin{equation} \label{eq:muscle_eq}
\tau = \sum_m j_m  f_m = \sum_m j_m  ( f_m^\mathrm{c} a + f_m^\mathrm{p}) = J^\mathrm{c} A + F^\mathrm{p},
\end{equation}
where $A=(a_0, a_1, ... , a_m)$ is a vector of muscle activation. 
The torque $\tau$ around a joint is the sum of muscle forces $f_m$ multiplied by its Jacobian $j_m$. The muscle force has two components: contractile $f_m^\mathrm{c}$ and passive $f_m^\mathrm{p}$. 
The contractile force is linearly proportional to the activation level $a$, while the passive force is produced when the muscle is stretched beyond a threshold. The state $s_\mathrm{joint}$ of muscles around a joint is represented by its force capacity $(F_\mathrm{min}^\mathrm{c}, F_\mathrm{max}^\mathrm{c}, F^\mathrm{p})$, where $F_\mathrm{min}^c$ and $F_\mathrm{max}^\mathrm{c}$, respectively, are the minimum and maximum contractile torque the muscles around a joint can produce and $F^\mathrm{p}$ is the aggregate passive force at the joint. We compute the minimum and maximum contractile torques approximately from the row-wise minimum and maximum values of the Jacobian matrix $J^\mathrm{c}$.

The gait state $s_\mathrm{gait}$ includes six terms $(\phi, \hat t, g_\mathrm{s}, g_\mathrm{c}, g_\mathrm{v}, h)$, where $\phi \in [0,1]$ is a phase of the gait cycle. The cycle begins with left heel strike at $\phi=0$, passes the halfway with right heel strike at $\phi=0.5$, and ends with the next left heel strike. The phase is different from the normalized time $\hat t \in [0,1]$ in the cycle, although they both parameterize the gait cycle. The normalized time increases at the fixed rate from zero to one and resets to zero repeatedly in the simulation, whereas the phase has variable step sizes to simulate (possibly asymmetric) gait variations. $g_\mathrm{s}$, $g_\mathrm{c}$, and $g_\mathrm{v}$ are the desired stride, cadence, and walking speed, respectively. The foothold location $h$ of the next heel strike can be derived from the desired stride.

Action $u=(\Delta M,\Delta \phi,\beta)$ includes pose displacement $\Delta M$, phase increment $\Delta \phi$, and threshold $\beta$. $\Delta M$ and $\Delta \phi$ describe how the reference gait $M_o(\phi)$ should be altered spatially and temporally to properly guide the simulation at state $s$. Specifically, the PD target 
\begin{equation} \label{eq:PDtarget}
    \hat M = M_o(\phi+\alpha \Delta \phi) \oplus \alpha \Delta M
\end{equation}
is a full-body pose of the skeletal system. Here, we use the motion displacement notation (see \cite{lee2008geometric} for detail). $\alpha \in [0,1]$ is the confidence level of the policy. With high confidence, the action is fully reflected in gait control. We will explain how to compute the confidence $\alpha$ and the threshold $\beta$ in the next section. The PD target thus obtained is fed into stable PD controller to convert it into the desired joint torque.

The phase is updated by $\Delta \phi$ at every simulation time step. The phase should be synchronized with the normalized time at the end of each gate cycle.  To do so, we use a clipping method during reinforcement learning. If the phase does not reach one yet at the end of the gait cycle ($\phi<1$ and $\hat t=1$), the phase jumps to one to synchronize. If the phase reaches one prematurely before the gait cycle ends ($\phi=1$ and $\hat t<1$), the phase is fixed to one until the end of the cycle. This simple clipping method is sufficient because the control policy gradually learns to synchronize smoothly in the process of reinforcement learning.

\subsection{Reward}

The reward $r$ includes four terms:
\begin{equation} \label{eq:reward}
r = r_\mathrm{head} r_\mathrm{stride} r_\mathrm{vel} + w_\mathrm{energy} r_\mathrm{energy}
\end{equation}
It is well-known that humans tend to stabilize their head during walking, because visual sensory and vestibular systems are in the head. $r_\mathrm{head}$ regularizes changes in linear velocity and rotation of the head:
\begin{equation} \label{eq:head_stability}
r_\mathrm{head} = \exp(- \frac{||\Delta v_\mathrm{head}||^2}{\sigma_{v}} - \frac{||\theta_\mathrm{head}||^2}{\sigma_{r}}),
\end{equation}
where $\Delta v_\mathrm{head}$ is the rate of change in linear velocity and $\theta_\mathrm{head}$ is the head orientation with respect to the frame align to the progression direction and the global up direction. 

$r_\mathrm{stride}$ and $r_\mathrm{vel}$, respectively, encourage to maintain the desired stride and walking speed. 
\begin{equation}
\begin{aligned}
r_\mathrm{stride} = \exp(- \frac{||h - h_\mathrm{desired}||^2}{\sigma_\mathrm{stride}}),\\ 
r_\mathrm{vel} = \exp(-\frac{||v - v_\mathrm{desired}||^2}{\sigma_\mathrm{vel}}).
\end{aligned}
\end{equation}
$r_\mathrm{energy}$ regularizes the energy expenditure during gait. 
\begin{equation} \label{eq:metabolic}
r_\mathrm{energy} = \exp(-\frac{||a||^2}{\sigma_\mathrm{energy}}),
\end{equation}
where $a$ is the muscle activation level. Note that we multiply the first three terms, but the last term is added to the others. This reward design is intended to enforce head stability, desired stride, and desired velocity simultaneously, since the multiplication gets rewarded only when all three terms get rewarded. The last term is intended to gently regularize excessive energy consumption by seeking a compromise between energy efficiency and gait requirements.

\subsection{Muscle Coordination}

Muscle coordination is the process of determining activation levels for all muscles that achieve desired joint torques. We employed the QP formulation for muscle coordination and its conversion to regression-by-supervised-learning presented by Lee et al~\cite{lee2019scalable}. Deriving from Eq (\ref{eq:muscle_eq}), the QP formulation is
\begin{equation} \label{eq:muscle-problem}
\begin{aligned}
\argmin_{A} ||\tau_\mathrm{desired} - (J^\mathrm{c} A + F^\mathrm{p} )||^2 + w_\mathrm{reg}||A||^2 \\
\text{subject to}\hspace{5mm} 0 \leq a_i \leq 1 \hspace{3mm}\text{for}\hspace{2mm} 0\le i \le m,
\end{aligned}
\end{equation}
The regression network $\pi_\psi$  learns the solution of the QP by supervised learning. Since DRL is episodic, it generates a large collection of simulation roll-outs in the learning phase. The regression network is jointly learned in the DRL framework by sampling tuples $(J^\mathrm{c},F^\mathrm{p},\tau)$ from the simulation roll-outs. Since $J^\mathrm{c}$ is a large, sparse matrix with $n$ columns and $m$ rows, where $n$ is the degrees of freedom of the skeleton and $m$ is the number of muscles, we vectorized the matrix for regression by compactly packing non-zero values. Then, the regression network learns the relation $\tau=\pi_\psi \big( {\rm vec}(J^\mathrm{c}),F^\mathrm{p} \big)$ from the tuples. The last layer of the regression network uses a sigmoid activation function to enforce the inequality condition $0 \leq a_i \leq 1$.

\subsection{Simulation and Learning}

In the learning phase, DRL picks random values for anatomy and gait conditions within a certain domain and runs physics-based simulation to evaluate its outcome repeatedly while gradually updating the policy and value networks. It is important to establish a stable initial configuration for each simulation. In particular, severe muscle contracture entails higher passive muscle force and a narrower joint ROM (Range of Motion). With low values for contracture conditions, skeletal poses selected from the reference gait may be outside the joint ROM. This violation can result in numerical instability of the simulation. To alleviate this risk, we set the initial configuration through pose optimization that minimizes
\begin{equation} \label{eq:pose-optimization}
\begin{aligned}
\min_{q} (F^\mathrm{p})^T W_\mathrm{joint} F^\mathrm{p} + w_\mathrm{com} || p_\mathrm{com} - p_\mathrm{StanceFoot} ||^2,
\end{aligned}
\end{equation}
where $q$ is a full-body pose of the skeletal system at the beginning of the simulation episode and $W_\mathrm{joint}$ is a diagonal weight matrix. The joints in the lower body are weighed more than the joints in the upper body since the lower body joints have a stronger influence on the gait. The first term penalizes the violation of the joint ROM, while the second term penalizes off-balance poses.

Although human gaits are not necessarily symmetric, the gait policy has an inherent symmetry because the control policy for left leg swing is also eligible for right leg swing if states and actions are properly reflected. We employed Phase-Based Mirroring~\cite{abdolhosseini2019learning} that allows us to learn the policy only for the half of the gait cycle $\phi\in[0,0.5]$. The policy of the other half $\phi\in[0.5,1]$ is the mirror reflection. This simple technique cuts the computation time in half and helps to produce symmetrical gait. Note that this symmetric policy can generate asymmetric gait without any difficulty.

\section{Cascaded Subsumption Learning} \label{sec:CSN}

The DRL algorithm in the previous section can successfully learn a parameterized control policy with a moderate number of adjustable parameters. This algorithm does not generalize easily to deal with more (100+) parameters even with carefully planned curriculum because the policy network easily forgets what it learned while learning a subsequent batch of curriculum. In this section, we present a new idea, cascaded subsumption, to learn a control policy with 618 adjustable parameters that include 10 body parameters $C_\mathrm{body}$ and 608 muscle parameters $C_\mathrm{muscle}=C_\mathrm{weakness} \cup C_\mathrm{contracture}$. The key to the scalability of Cascaded Subsumption Network (CSN) is its ability to learn new knowledge progressively while keeping previously learned knowledge unaltered by over-learning.   

Let $\{c^j_i\}$ be a partition of the muscle parameters at level $0<j< N$.
\begin{equation}
    C_\mathrm{muscle} = \bigcup_i c^j_i \;\;\text{for any}\; j.
\end{equation}
The multiple subsets at level $j-1$ are subsumed into a parent set at level $j$. The child sets are a partition of their parent set. Each subset can have only one parent except for the ones at the base and top layers.

The CSN consists of $N$ network layers $\{\pi_i^j\}$ for $0\le j< N$. There is only one network in the base level $0$ and the top level $N-1$. Each network layer is associated with a subset of muscle parameters except for $\pi^0$ at the base level, which does not take any adjustable muscle parameters. The networks are learned progressively level-by-level. The base network $\pi^0$ learns a control policy of healthy (all weakness and contracture parameters are fixed to 1) individuals parameterized only by body and gait parameters. 
The network $\pi_i^1$ at the next layer learns how muscle disorders associated with $c_i^1$ affect the gait. The new knowledge $\pi_i^1$ learned over the base network $\pi^0$ is reflected to gait control by overlaying actions. Specifically, we add the spatial and temporal displacements of the two networks to compute the PD target for $\pi_i^1$. Eq (\ref{eq:PDtarget}) is replaced by
\begin{equation}
    \hat M = M_o(\phi+ \Delta \phi^0 + \alpha^1_i \Delta \phi^1_i) \oplus \Delta M^0 \oplus \alpha^1_i \Delta M^1_i,
\end{equation}
where $(\Delta \phi^0, \Delta M^0)$ and $(\Delta \phi^1_i, \Delta M^1_i)$ are part of action of $\pi^0$ and $\pi_i^1$, respectively. While $\pi^0$ is always confident about its action, the confidence of $\pi_i^1$ depends on state $s$. For example, if the muscles in state $s$ are all healthy and normal, $\pi^0$ already knows how to control and simulation the gait. The intervention of $\pi_i^1$ risks over-learning and forgetting, so the confidence $\alpha^1_i$ should be low. If some muscles in $c^1_i$ has weakness or contracture, $\pi_i^1$ should have higher confidence to exploit new knowledge it gained.

\begin{algorithm}
\fontsize{9pt}{10pt}\selectfont
\begin{algorithmic}
\caption{Confidence calculation}\label{alg:confidence}
\Statex $\pi^n$ : a network at level $n$
\Statex $\pi^{n-1}_1, \pi^{n-1}_2, \cdots, \pi^{n-1}_m$ : Child networks of $\pi^n$.
\Statex $\beta$ : threshold value learned by $\pi^n (u|s)$
\Statex $\alpha(u,s)$ : confidence of taking action $u$ given state $s$
\Statex $W_\mathrm{joint}$ : a diagonal weight matrix of joints
\State
\For{$i = 1, \cdots , m$}
    \State $\Delta c_i \gets c^n \setminus c_i^{n-1}$ 
    \State $\Delta s_\mathrm{joint} \gets \text{Changes in joint state} \;\;  s_\mathrm{joint} \;\; \text{induced by}\;\; \Delta c_i$
    \State $d \gets \|(\Delta s_\mathrm{joint})^\top W_\mathrm{joint} \Delta s_\mathrm{joint} \|$
    \If {$d_\mathrm{min} < d$}
        {$d_\mathrm{min}  \gets  d$}
    \EndIf
\EndFor
\State $\alpha \gets \mathrm{sigmoid}(d_\mathrm{min} - \beta)$
\end{algorithmic}
\end{algorithm}

The general procedure of computing the confidence is as follows (see Algorithm~\ref{alg:confidence}). Let $\pi^n$ by any network at level $n$ and $\pi^{n-1}_1, \pi^{n-1}_2, \cdots, \pi^{n-1}_m$ be its child networks at level $n-1$. $\pi^n$ is a super-set of any of its children. We evaluate the representation power of each network based on its force capacity around joints, which is described by $s_\mathrm{joint}=(F_\mathrm{min}^\mathrm{c}, F_\mathrm{max}^\mathrm{c}, F^\mathrm{p})$. If inclusion of new muscle parameters $\Delta c_i$ makes impact on its force capacity, $\pi^n$ has to learn new knowledge over its child networks and thus has high confidence. The threshold $\beta$ decides how big the impact should be to have confidence. $\beta$ is part of policy action learned in RL.

Muscle coordination also requires cascaded subsumption. Composition of muscle regression networks are different from composition of policy networks since muscle activation is bounded between zero and one. There is no guarantee that the sum of muscle activation outputs weighed by confidence is within $[0,1]$. To alleviate this problem, we take muscle activation output $A_\mathrm{unnormalized}$ before it passes through the sigmoid function, weighted sum the outputs of the overlaying networks, and feed it back to the sigmoid activation function. This simple technique works well with cascaded regression networks.

\begin{figure}
    \centering
    \includegraphics[width=\linewidth]{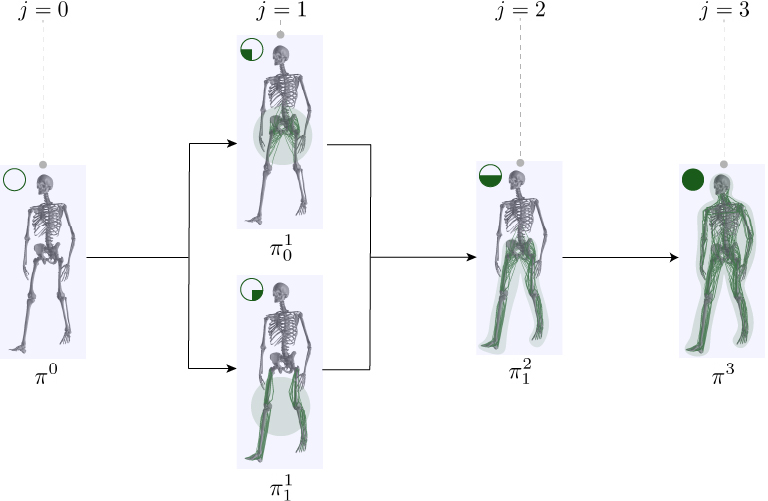}
    \caption{\label{fig:CSN_Connectivity}. An example of cascaded subsumption}
\end{figure}

The first version of our Generative GaitNet has four CSN layers (see Figure~\ref{fig:CSN_Connectivity}). The base network learned the influence of body and gait parameters. There are two networks at level 1. One of them learned the influence of muscle disorders around hip joints, while the other network learned the muscle parameters around knee and ankle joints. The network at level 2 subsumes its two child networks to learn control policies with an entire set of lower body muscles. Since the influences of upper body muscles are relatively weaker, we did not explicitly construct networks associated with the upper body at level 1 and 2. The last network at level 3 managed to learn the co-occurrence relation between upper and lower body muscles without upper body child networks.

\section{Experiments and Result} 

Our simulation and learning system is implemented in C++ and Python. The skeletal system and Hill-type muscle contracture dynamics are implemented on the DART physics engine~\cite{lee2018dart}. The skeleton consists of 23 rigid bones connected by 8 revolute joints at elbows, knees, and toes and 14 ball-and-socket joints. In total, the skeletal system has 50 degrees of freedom. The physics simulation time step is 480 Hz, the PD control time step is 30 Hz, and the muscle activation time step is 480 Hz. Deep reinforcement learning uses a PPO (Proximal Policy Optimization) algorithm with adaptive KL penalty and clipping implemented in Ray RLlib. Each policy/value network consists of three fully connected layers of 512 nodes and each value function, while each muscle network consists of three fully connected layers of 256 nodes. The learning parameters are summarized in table~\ref{table:learning_parameter}.

\begin{table}
\caption{\label{table:learning_parameter} learning parameters}
\begin{center}
\begin{tabular}{|l|c|}
\hline
{Policy/Value and Muscle learning rate} & {$1.0e^{-4}$}\\
\hline
{Discount factor ($\gamma$)} & {0.99}\\
\hline
{GAE and TD ($\lambda$)} & {0.99}\\
\hline
{\# of tuples per policy update} & {65536}\\
\hline
{Batch size for policy update} & {1024}\\
\hline
{Iteration for policy/value update} & {4}\\
\hline
{Iteration for muscle update} & {4-10}\\
\hline
{Maximum time horizon (sec)} & {10}\\
\hline
{Clip parameter ($\epsilon$)} & {0.2}\\
\hline
{KL target ($\epsilon$)} & {0.01}\\
\hline
{Minimum tuples per selected parameter} & {460}\\
\hline
\end{tabular}
\end{center}
\end{table}

The parameter domain is spanned by body, muscle, gait parameters. We can adjust the body and head size by 15\% and the limb length by 10\% in the domain. Stride and cadence ranges from $-25\%$ to $+25\%$. Ideally, both weakness and contracture parameters range from zero to one. In practice, low values for weakness parameters close to zero can have singularity issues. Therefore, their lower bound are set greater than zero. We determined the lower bound of contracture parameters based on the magnitude of passive force incurred by contraction at a neutral pose. The lower bound ranges from 0.5 to 0.8.

\begin{figure}
    \centering
    \includegraphics[width=1.0\linewidth]{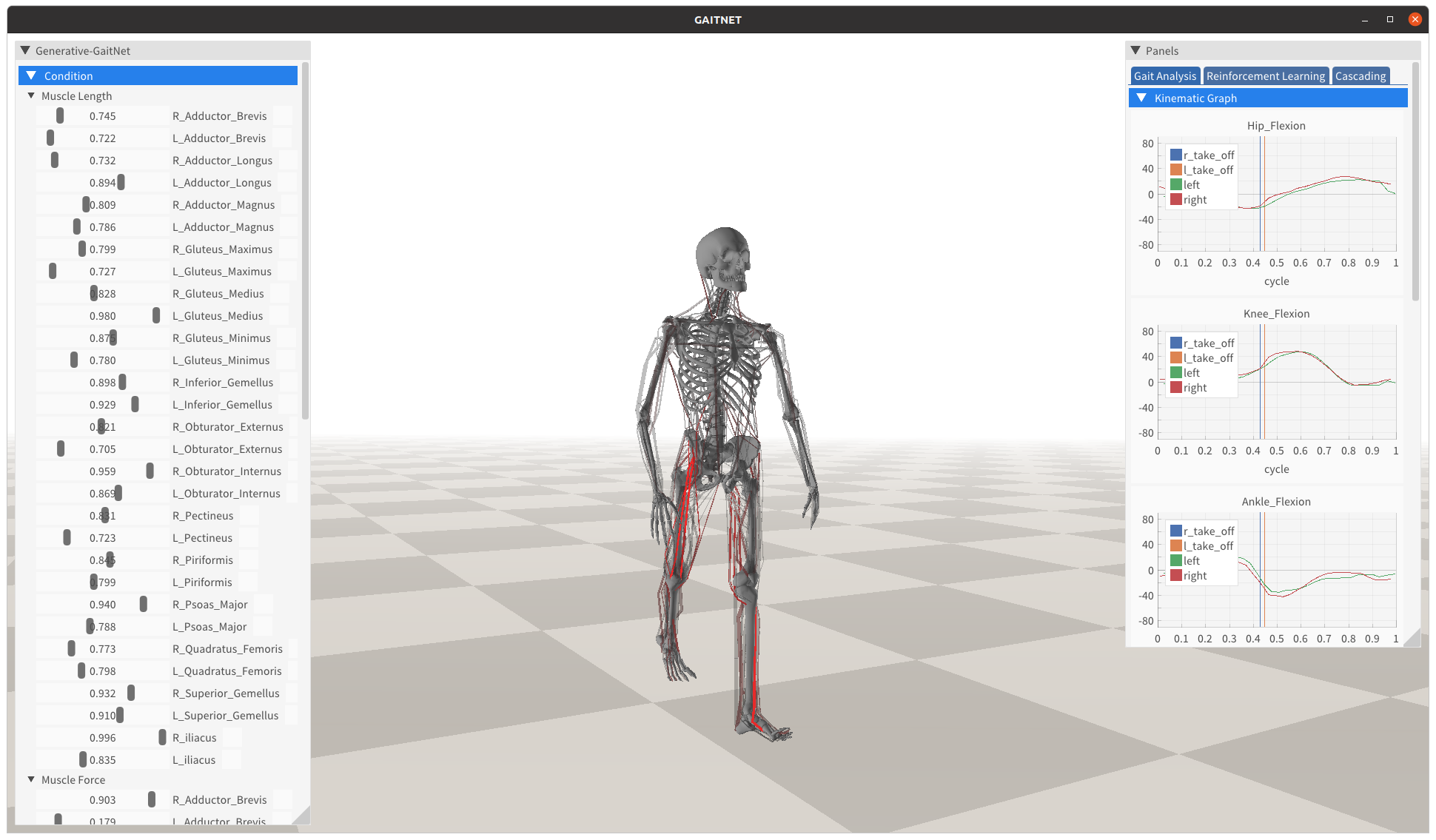}
    \caption{\label{fig:interface}. Interactive user interface for real-time gait simulation}
\end{figure}

We used a cluster server with four Intel Xeon 6242 CPU 2.8GHz and a Nvidia RTX 3090 to learn the GaitNet. Learning a single network in CSN requires approximately 100 million simulation roll-outs and takes 6 hours to 24 hours in the cluster. Learning all layers of CSN takes 4 to 5 days.
Once the GaitNet is learned, gait simulation runs in real-time. Through our interactive user interface system, the user can change the body size and proportion, limb length, stride, cadence, and any of muscle parameters interactively (see Figure~\ref{fig:interface}). The change is reflected immediately in real-time simulation. We will release the code and the pretrained network in public repository.

\subsection{Pathological Gaits}

We reproduced many pathological gaits with Generative GaitNet to demonstrate its expressive power. We manually tuned body and muscle parameters to match the gaits in video clips. The simulation results are best viewed in the supplementary video.

\subsubsection{Foot Drop.}
Foot drop is a condition in which someone has difficulty lifting the front part of the foot. To simulate foot drop, we made muscles around the ankle weak (weakness parameter = 0.05). The ankle dorsiflexor muscles include tibialis anterior, extensor hallucis longus, and extensor digitorum longus. The MASS system is probably the current state-of-the-art in musculoskeletal simulation~\cite{lee2019scalable}. The MASS system equipped with imitation rewards demonstrated excellent quality and performance in simulating pathological gaits induced by muscle contraction, which is enforced by inequality constraints in simulation. However, the reference-tracking method is not adept at dealing with muscle weakness probably because DRL is too smart. Even with weak muscles, it often manages to find a clever way to imitate the reference trajectory and this overly clever way does not look human-like. Our GaitNet is not hindered by imitation requirements and thus has a better chance to simulate both weakness and contracture accurately.

\subsubsection{Equinus.}
Equinus is a condition in which the upward bending motion (dorsiflexion) of the ankle is limited. Someone with equinus tends to walk on their toes. We tightened ankle plantarflexor muscles (gastrocnemius and soleus) to simulate toe walking (contracture 0.8). 

\subsubsection{Lumbar Hyperlordosis.}
Psoas major muscles are power hip flexors. Contracture (0.65) in psoas major muscles lifts the upper leg towards the body and consequently makes someone with this condition bend forward during walking. There is a secondary effect of compensating for forward bending, called lumbar hyperlordosis. Lordosis is the inward curve of the lumbar spine. Hyperlordosis often occurs to compensate for tight psoas muscles. The MASS system can only simulate the first effect (forward bending), while our GaitNet predicts the subsequent effect more accurately to produce lumbar hyperlordosis.

\subsubsection{Stiff Knee.}
We reproduced stiff knee gaits by setting contracture in vastus muscles (0.87), tibialis posterior (0.78), peroneus tertius (0.71), gluteus muscles (0.85), and extensor digiti mini (0.87).

\subsubsection{Crouch Gait.}
Crouch gait is characterized by knee flexion during the stance phase. This gait disorder is common among patients with cerebral palsy. We reproduced the crouch gait by setting contracture in adductor brevis (0.7), psoas major (0.65), bicep femoris short (0.65), extensor digitorum longus (0.8) extensor hallucis longus (0.75), flexor digiti minimi brevis (0.78), tibialis anterior (0.8), and tibialis posterior (0.78).

\begin{figure}
    \includegraphics[width=\linewidth]{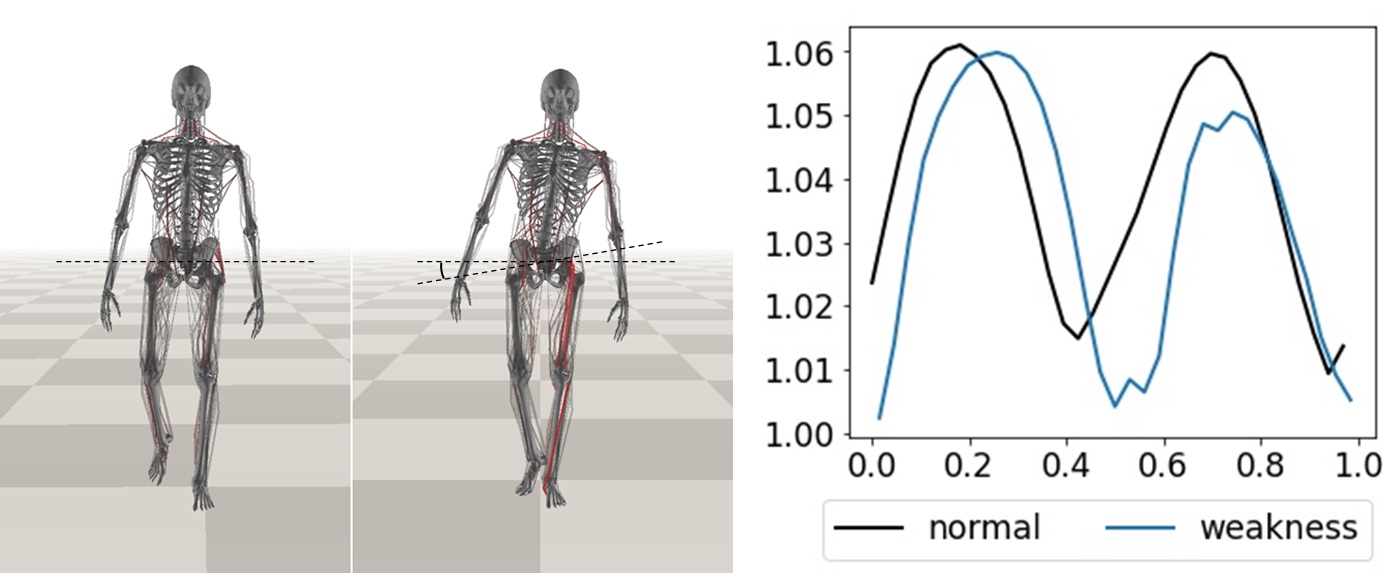}
    \caption{\label{fig:trendelenburg} Normal gait and Trendelenburg gait. Pelvic obliquity plots in the gait cycle.}
\end{figure}

\subsubsection{Trendelenburg and Waddling Gait.}
Trendelenburg gait is an abnormal gait caused by weakness of gluteal muscles. In normal gait, the pelvis is balanced by the gluteal muscles. The weakness of gluteal muscles on one side results in pelvic drop on the opposite side and the upper body leans toward the weakened side to compensate for the pelvic drop (see Figure~\ref{fig:trendelenburg}). The GaitNet successfully produces both pelvic drop and upper body leaning. Gluteal weakness on both sides causes waddling gait, which is characterized by upper body swaying.

\begin{figure}
    \includegraphics[width=\linewidth]{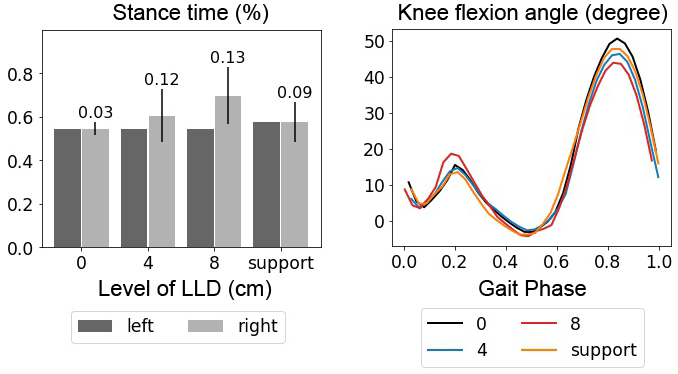}
    \caption{\label{fig:LLD} Leg length discrepancy. }
\end{figure}

\subsubsection{Leg Length Discrepansy}
Leg length discrepansy is a condition in which one leg is shorter than the other (see Figure~\ref{fig:LLD}). It has been reported that the longer leg has a longer stance duration than the shorter leg. Bhave et al~\cite{bhave1999improvement} showed this trend in their experiments with human subjects. We reproduced their experiments in physics-based simulation with GaitNet. We generated four cases. The first case has an even leg length. The second case has one leg 4cm shorter than the other. The third case has one leg 8cm shorter than the other. The fourth case uses a thick insole to compensate for 8cm discrepansy in leg length. We found that the simulation results closely matched the human experiments.

\subsection{Comparison to Curriculum Learning}
The biggest advantage of CSN is its ability to keep the previously learned policy unaltered while learning policies in a new parameter domain. The normal gait in a healthy condition generated by the base network $\pi^0$ is almost identical to the gait in the same condition generated by the top-layer network $\pi^3$. It is because the knowledge learned by the base network remains intact while learning networks in the subsequent layers. Brute-force curriculum learning with a single (non-layered) network does not guarantee this property. If the network learns the normal gait first and then explores in the parameter domain according to curriculum, the normal gait the network remembers would change as the learning proceeds because the new knowledge interferes the previous learned knowledge.

\section{Discussion}

Generative GaitNet successfully learned the relation between anatomy and gait of the human musculoskeletal system. With the pre-trained GaitNet, predictive gait simulation can be a practical tool for physicians, clinical practitioners, and biomechanics/ergonomics/sports researchers seeking to understand how anatomical conditions affect human gait. We anticipate that Generative GaitNet will be used to design wearable walking assist devices and predict the outcome of orthopedic surgical procedures for improving pathological gaits.

Cascaded Subsumption Networks are remarkably adept at dealing with the high-dimensional parameter domain. With 618 adjustable parameters, simply examining all corners of the domain would require a tremendously large number ($2^{618} \simeq 10^{186}$) of simulation roll-outs. We learned the first version of our GaitNet with only 340 million simulation roll-outs, which is computationally demanding but manageable in modern CPU/GPU clusters. This result shows the scalability and efficacy of our approach.

Generative GaitNet also has numerous limitations. Our musculoskeletal model lacks many anatomical features such as nervous system, ligaments, skin and soft tissue. The nervous system is particularly important because it is an integral part of the sensory-motor system. It would be possible to simulate Parkinson gait and spasticity in cerebral palsy patients. Volumetric FEM modeling of muscles offers promising opportunities to achieve more accurate simulations of muscle contraction dynamics~\cite{lee2018dexterous,abdrashitov2021interactive}.


Although Generative GaitNet currently focuses on simulating bipedal walking, there are numerous possibilities to generalize its scope because the learning procedure is based on fundamental principles, such as energy minimization and head stabilization. These principles are also applicable to any type of locomotion such as bipedal running, quadrupedal locomotion, swimming and even flapping flying~\cite{yu2018learning, ju2013data}.


\bibliographystyle{spmpsci}
\bibliography{reference} 

\begin{thebibliography}{10}
\providecommand{\url}[1]{{#1}}
\providecommand{\urlprefix}{URL }
\expandafter\ifx\csname urlstyle\endcsname\relax
  \providecommand{\doi}[1]{DOI~\discretionary{}{}{}#1}\else
  \providecommand{\doi}{DOI~\discretionary{}{}{}\begingroup
  \urlstyle{rm}\Url}\fi

\bibitem{abdolhosseini2019learning}
Abdolhosseini, F., Ling, H.Y., Xie, Z., Peng, X.B., van~de Panne, M.: On
  learning symmetric locomotion.
\newblock In: Motion, Interaction and Games, pp. 1--10 (2019)

\bibitem{abdrashitov2021interactive}
Abdrashitov, R., Bang, S., Levin, D.I., Singh, K., Jacobson, A.: Interactive
  modelling of volumetric musculoskeletal anatomy.
\newblock arXiv preprint arXiv:2106.05161  (2021)

\bibitem{al2012trajectory}
Al~Borno, M., De~Lasa, M., Hertzmann, A.: Trajectory optimization for full-body
  movements with complex contacts.
\newblock IEEE transactions on visualization and computer graphics
  \textbf{19}(8), 1405--1414 (2012)

\bibitem{anand2019deep}
Anand, A.S., Zhao, G., Roth, H., Seyfarth, A.: A deep reinforcement learning
  based approach towards generating human walking behavior with a neuromuscular
  model.
\newblock In: 2019 IEEE-RAS 19th International Conference on Humanoid Robots
  (Humanoids), pp. 537--543. IEEE (2019)

\bibitem{anderson2001dynamic}
Anderson, F.C., Pandy, M.G.: Dynamic optimization of human walking.
\newblock J. Biomech. Eng. \textbf{123}(5), 381--390 (2001)

\bibitem{bergamin2019drecon}
Bergamin, K., Clavet, S., Holden, D., Forbes, J.R.: Drecon: data-driven
  responsive control of physics-based characters.
\newblock ACM Transactions On Graphics (TOG) \textbf{38}(6), 1--11 (2019)

\bibitem{bhave1999improvement}
Bhave, A., Paley, D., Herzenberg, J.E.: Improvement in gait parameters after
  lengthening for the treatment of limb-length discrepancy.
\newblock JBJS \textbf{81}(4), 529--34 (1999)

\bibitem{coros2010generalized}
Coros, S., Beaudoin, P., Van~de Panne, M.: Generalized biped walking control.
\newblock ACM Transactions On Graphics (TOG) \textbf{29}(4), 1--9 (2010)

\bibitem{delp2007opensim}
Delp, S.L., Anderson, F.C., Arnold, A.S., Loan, P., Habib, A., John, C.T.,
  Guendelman, E., Thelen, D.G.: Opensim: open-source software to create and
  analyze dynamic simulations of movement.
\newblock IEEE transactions on biomedical engineering \textbf{54}(11),
  1940--1950 (2007)

\bibitem{dembia2020opensim}
Dembia, C.L., Bianco, N.A., Falisse, A., Hicks, J.L., Delp, S.L.: Opensim moco:
  musculoskeletal optimal control.
\newblock PLOS Computational Biology \textbf{16}(12), e1008,493 (2020)

\bibitem{falisse2019rapid}
Falisse, A., Serrancol{\'\i}, G., Dembia, C.L., Gillis, J., Jonkers, I.,
  De~Groote, F.: Rapid predictive simulations with complex musculoskeletal
  models suggest that diverse healthy and pathological human gaits can emerge
  from similar control strategies.
\newblock Journal of the Royal Society Interface \textbf{16}(157), 20190,402
  (2019)

\bibitem{fussell2021supertrack}
Fussell, L., Bergamin, K., Holden, D.: Supertrack: motion tracking for
  physically simulated characters using supervised learning.
\newblock ACM Transactions on Graphics (TOG) \textbf{40}(6), 1--13 (2021)

\bibitem{geijtenbeek2013flexible}
Geijtenbeek, T., Van De~Panne, M., Van Der~Stappen, A.F.: Flexible muscle-based
  locomotion for bipedal creatures.
\newblock ACM Transactions on Graphics (TOG) \textbf{32}(6), 1--11 (2013)

\bibitem{heess2017emergence}
Heess, N., TB, D., Sriram, S., Lemmon, J., Merel, J., Wayne, G., Tassa, Y.,
  Erez, T., Wang, Z., Eslami, S., et~al.: Emergence of locomotion behaviours in
  rich environments.
\newblock arXiv preprint arXiv:1707.02286  (2017)

\bibitem{hodgins1995animating}
Hodgins, J.K., Wooten, W.L., Brogan, D.C., O'Brien, J.F.: Animating human
  athletics.
\newblock In: Proceedings of the 22nd annual conference on Computer graphics
  and interactive techniques, pp. 71--78 (1995)

\bibitem{jiang2019synthesis}
Jiang, Y., Van~Wouwe, T., De~Groote, F., Liu, C.K.: Synthesis of biologically
  realistic human motion using joint torque actuation.
\newblock ACM Transactions On Graphics (TOG) \textbf{38}(4), 1--12 (2019)

\bibitem{ju2013data}
Ju, E., Won, J., Lee, J., Choi, B., Noh, J., Choi, M.G.: Data-driven control of
  flapping flight.
\newblock ACM Transactions on Graphics (TOG) \textbf{32}(5), 1--12 (2013)

\bibitem{kajita2003biped}
Kajita, S., Kanehiro, F., Kaneko, K., Fujiwara, K., Harada, K., Yokoi, K.,
  Hirukawa, H.: Biped walking pattern generation by using preview control of
  zero-moment point.
\newblock In: 2003 IEEE International Conference on Robotics and Automation
  (Cat. No. 03CH37422), vol.~2, pp. 1620--1626. IEEE (2003)

\bibitem{kidzinski2018learning}
Kidzi{\'n}ski, {\L}., Mohanty, S.P., Ong, C.F., Huang, Z., Zhou, S., Pechenko,
  A., Stelmaszczyk, A., Jarosik, P., Pavlov, M., Kolesnikov, S., et~al.:
  Learning to run challenge solutions: Adapting reinforcement learning methods
  for neuromusculoskeletal environments.
\newblock In: The NIPS'17 Competition: Building Intelligent Systems, pp.
  121--153. Springer (2018)

\bibitem{kidzinski2020artificial}
Kidzi{\'n}ski, {\L}., Ong, C., Mohanty, S.P., Hicks, J., Carroll, S., Zhou, B.,
  Zeng, H., Wang, F., Lian, R., Tian, H., et~al.: Artificial intelligence for
  prosthetics: Challenge solutions.
\newblock In: The NeurIPS'18 Competition, pp. 69--128. Springer (2020)

\bibitem{kwon2017momentum}
Kwon, T., Hodgins, J.K.: Momentum-mapped inverted pendulum models for
  controlling dynamic human motions.
\newblock ACM Transactions on Graphics (TOG) \textbf{36}(1), 1--14 (2017)

\bibitem{kwon2020fast}
Kwon, T., Lee, Y., Van De~Panne, M.: Fast and flexible multilegged locomotion
  using learned centroidal dynamics.
\newblock ACM Transactions on Graphics (TOG) \textbf{39}(4), 46--1 (2020)

\bibitem{lee2008geometric}
Lee, J.: Representing rotations and orientations in geometric computing.
\newblock IEEE Computer Graphics and Applications \textbf{28}(2), 75--83 (2008)

\bibitem{lee2018dart}
Lee, J., Grey, M.X., Ha, S., Kunz, T., Jain, S., Ye, Y., Srinivasa, S.S.,
  Stilman, M., Liu, C.K.: Dart: Dynamic animation and robotics toolkit.
\newblock The Journal of Open Source Software \textbf{3}(22), 500 (2018)

\bibitem{lee2021learning}
Lee, S., Lee, S., Lee, Y., Lee, J.: Learning a family of motor skills from a
  single motion clip.
\newblock ACM Transactions on Graphics (TOG) \textbf{40}(4), 1--13 (2021)

\bibitem{lee2019scalable}
Lee, S., Park, M., Lee, K., Lee, J.: Scalable muscle-actuated human simulation
  and control.
\newblock ACM Transactions On Graphics (TOG) \textbf{38}(4), 1--13 (2019)

\bibitem{lee2018dexterous}
Lee, S., Yu, R., Park, J., Aanjaneya, M., Sifakis, E., Lee, J.: Dexterous
  manipulation and control with volumetric muscles.
\newblock ACM Transactions on Graphics (TOG) \textbf{37}(4), 1--13 (2018)

\bibitem{lee2010data}
Lee, Y., Kim, S., Lee, J.: Data-driven biped control.
\newblock In: ACM SIGGRAPH 2010 papers, pp. 1--8 (2010)

\bibitem{lee2014locomotion}
Lee, Y., Park, M.S., Kwon, T., Lee, J.: Locomotion control for many-muscle
  humanoids.
\newblock ACM Transactions on Graphics (TOG) \textbf{33}(6), 1--11 (2014)

\bibitem{liu2018learning}
Liu, L., Hodgins, J.: Learning basketball dribbling skills using trajectory
  optimization and deep reinforcement learning.
\newblock ACM Transactions on Graphics (TOG) \textbf{37}(4), 1--14 (2018)

\bibitem{merel2020catch}
Merel, J., Tunyasuvunakool, S., Ahuja, A., Tassa, Y., Hasenclever, L., Pham,
  V., Erez, T., Wayne, G., Heess, N.: Catch \& carry: reusable neural
  controllers for vision-guided whole-body tasks.
\newblock ACM Transactions on Graphics (TOG) \textbf{39}(4), 39--1 (2020)

\bibitem{min2019softcon}
Min, S., Won, J., Lee, S., Park, J., Lee, J.: Softcon: Simulation and control
  of soft-bodied animals with biomimetic actuators.
\newblock ACM Transactions on Graphics (TOG) \textbf{38}(6), 1--12 (2019)

\bibitem{mordatch2010robust}
Mordatch, I., De~Lasa, M., Hertzmann, A.: Robust physics-based locomotion using
  low-dimensional planning.
\newblock In: ACM SIGGRAPH 2010 papers, pp. 1--8 (2010)

\bibitem{mordatch2014combining}
Mordatch, I., Todorov, E.: Combining the benefits of function approximation and
  trajectory optimization.
\newblock In: Robotics: Science and Systems, vol.~4 (2014)

\bibitem{ong2019predicting}
Ong, C.F., Geijtenbeek, T., Hicks, J.L., Delp, S.L.: Predicting gait
  adaptations due to ankle plantarflexor muscle weakness and contracture using
  physics-based musculoskeletal simulations.
\newblock PLoS computational biology \textbf{15}(10), e1006,993 (2019)

\bibitem{park2019learning}
Park, S., Ryu, H., Lee, S., Lee, S., Lee, J.: Learning predict-and-simulate
  policies from unorganized human motion data.
\newblock ACM Transactions on Graphics (TOG) \textbf{38}(6), 1--11 (2019)

\bibitem{peng2018deepmimic}
Peng, X.B., Abbeel, P., Levine, S., van~de Panne, M.: Deepmimic: Example-guided
  deep reinforcement learning of physics-based character skills.
\newblock ACM Transactions on Graphics (TOG) \textbf{37}(4), 1--14 (2018)

\bibitem{peng2016terrain}
Peng, X.B., Berseth, G., Van~de Panne, M.: Terrain-adaptive locomotion skills
  using deep reinforcement learning.
\newblock ACM Transactions on Graphics (TOG) \textbf{35}(4), 1--12 (2016)

\bibitem{peng2017deeploco}
Peng, X.B., Berseth, G., Yin, K., Van De~Panne, M.: Deeploco: Dynamic
  locomotion skills using hierarchical deep reinforcement learning.
\newblock ACM Transactions on Graphics (TOG) \textbf{36}(4), 1--13 (2017)

\bibitem{peng2017learning}
Peng, X.B., van~de Panne, M.: Learning locomotion skills using deeprl: Does the
  choice of action space matter?
\newblock In: Proceedings of the ACM SIGGRAPH/Eurographics Symposium on
  Computer Animation, pp. 1--13 (2017)

\bibitem{portelas2020teacher}
Portelas, R., Colas, C., Hofmann, K., Oudeyer, P.Y.: Teacher algorithms for
  curriculum learning of deep rl in continuously parameterized environments.
\newblock In: Conference on Robot Learning, pp. 835--853. PMLR (2020)

\bibitem{ryu2021functionality}
Ryu, H., Kim, M., Lee, S., Park, M.S., Lee, K., Lee, J.: Functionality-driven
  musculature retargeting.
\newblock In: Computer Graphics Forum, vol.~40, pp. 341--356. Wiley Online
  Library (2021)

\bibitem{sok2007simulating}
Sok, K.W., Kim, M., Lee, J.: Simulating biped behaviors from human motion data.
\newblock In: ACM SIGGRAPH 2007 papers, pp. 107--es (2007)

\bibitem{song2018predictive}
Song, S., Geyer, H.: Predictive neuromechanical simulations indicate why
  walking performance declines with ageing.
\newblock The Journal of physiology \textbf{596}(7), 1199--1210 (2018)

\bibitem{song2020deep}
Song, S., Kidzi{\'n}ski, {\L}., Peng, X.B., Ong, C., Hicks, J.L., Levine, S.,
  Atkeson, C., Delp, S.: Deep reinforcement learning for modeling human
  locomotion control in neuromechanical simulation.
\newblock bioRxiv  (2020)

\bibitem{tassa2012synthesis}
Tassa, Y., Erez, T., Todorov, E.: Synthesis and stabilization of complex
  behaviors through online trajectory optimization.
\newblock In: 2012 IEEE/RSJ International Conference on Intelligent Robots and
  Systems, pp. 4906--4913. IEEE (2012)

\bibitem{thatte2015toward}
Thatte, N., Geyer, H.: Toward balance recovery with leg prostheses using
  neuromuscular model control.
\newblock IEEE Transactions on Biomedical Engineering \textbf{63}(5), 904--913
  (2015)

\bibitem{wang2019terrain}
Wang, J., Qin, W., Sun, L.: Terrain adaptive walking of biped neuromuscular
  virtual human using deep reinforcement learning.
\newblock IEEE Access \textbf{7}, 92,465--92,475 (2019)

\bibitem{wang2009optimizing}
Wang, J.M., Fleet, D.J., Hertzmann, A.: Optimizing walking controllers.
\newblock In: ACM SIGGRAPH Asia 2009 papers, pp. 1--8 (2009)

\bibitem{wang2012optimizing}
Wang, J.M., Hamner, S.R., Delp, S.L., Koltun, V.: Optimizing locomotion
  controllers using biologically-based actuators and objectives.
\newblock ACM Transactions on Graphics (TOG) \textbf{31}(4), 1--11 (2012)

\bibitem{waterval2021validation}
Waterval, N., Veerkamp, K., Geijtenbeek, T., Harlaar, J., Nollet, F., Brehm,
  M., van~der Krogt, M.: Validation of forward simulations to predict the
  effects of bilateral plantarflexor weakness on gait.
\newblock Gait \& Posture \textbf{87}, 33--42 (2021)

\bibitem{won2020scalable}
Won, J., Gopinath, D., Hodgins, J.: A scalable approach to control diverse
  behaviors for physically simulated characters.
\newblock ACM Transactions on Graphics (TOG) \textbf{39}(4), 33--1 (2020)

\bibitem{won2019learning}
Won, J., Lee, J.: Learning body shape variation in physics-based characters.
\newblock ACM Transactions on Graphics (TOG) \textbf{38}(6), 1--12 (2019)

\bibitem{yin2008continuation}
Yin, K., Coros, S., Beaudoin, P., Van~de Panne, M.: Continuation methods for
  adapting simulated skills.
\newblock In: ACM SIGGRAPH 2008 papers, pp. 1--7 (2008)

\bibitem{yin2007simbicon}
Yin, K., Loken, K., Van~de Panne, M.: Simbicon: Simple biped locomotion
  control.
\newblock ACM Transactions on Graphics (TOG) \textbf{26}(3), 105--es (2007)

\bibitem{yu2018learning}
Yu, W., Turk, G., Liu, C.K.: Learning symmetric and low-energy locomotion.
\newblock ACM Transactions on Graphics (TOG) \textbf{37}(4), 1--12 (2018)

\end{thebibliography}

\end{document}